\documentclass[aps,twocolumn,pra,groupedaddress,superscriptaddress,nofootinbib]{revtex4}
\usepackage{graphicx,amsmath,amssymb,amsbsy,subfigure,hyperref,bbm,times,txfonts}

\hypersetup{
   colorlinks=true,
   linkcolor=blue,
}
\graphicspath{{figure/}}
\begin{document}

\newcommand{\ket}[1]{|#1\rangle}
\newcommand{\bra}[1]{\langle#1|}
\newcommand{\ketbra}[1]{| #1\rangle\!\langle #1 |}
\newcommand{\kebra}[2]{| #1\rangle\!\langle #2 |}
\newcommand{\id}{\mathbbm{1}}
\newcommand{\ohm}{\Omega_{\rm CQ}}
\newcommand{\rhobd}{\rho^{\vec{c}}_{AB}}
\providecommand{\tr}[1]{\text{tr}\left[#1\right]}
\providecommand{\tra}[1]{\text{tr}_A\left[#1\right]}
\providecommand{\trb}[1]{\text{tr}_B\left[#1\right]}
\providecommand{\abs}[1]{\left|#1\right|}
\providecommand{\sprod}[2]{\langle#1|#2\rangle}
\providecommand{\expect}[2]{\bra{#2} #1 \ket{#2}}

\title{Entanglement transfer in a noisy cavity network with parity-deformed radiation fields}

\author{Alireza Dehghani}
\email{alireza.dehghani@gmail.com}
\affiliation{Department of Physics, Payame Noor University, P.O.Box 19395-3697, Tehran, Iran}

\author{Bashir Mojaveri}
\email{bmojaveri@azaruniv.ac.ir}
\affiliation{Department of Physics, Azarbaijan Shahid Madani University, P.O.Box 51745-406, Tabriz, Iran}

\author{Rasoul Jafarzadeh Bahrbeig}
\email{r.jafarzadeh@azaruniv.ac.ir}
\affiliation{Department of Physics, Azarbaijan Shahid Madani University, P.O.Box 51745-406, Tabriz, Iran}

\author{Farzam Nosrati}
%\email{farzam.nosraty@gmail.com}
\affiliation{Dipartimento di Ingegneria, Universit\`{a} di Palermo, Viale delle Scienze, Edificio 9, 90128 Palermo, Italy}
\affiliation{INRS-EMT, 1650 Boulevard Lionel-Boulet, Varennes, Qu\'{e}bec J3X 1S2, Canada}

\author{Rosario Lo Franco}
\email{rosario.lofranco@unipa.it}
\affiliation{Dipartimento di Ingegneria, Universit\`{a} di Palermo, Viale delle Scienze, Edificio 6, 90128 Palermo, Italy}
\affiliation{Dipartimento di Fisica e Chimica, Universit\`a di Palermo, via Archirafi 36, 90123 Palermo, Italy}

\begin{abstract}
We investigate the effects of parity-deformed radiation fields on the dynamics of entanglement transfer to distant noninteracting atom qubits. These qubits are embedded in two separated lossy cavities connected by a leaky fiber, which acts as a cavity buffer with delocalized modes.  
The process is studied within a single-excitation subspace, the parity-deformed cavity photons allowing the introduction of static local classical fields which function as a control. The mechanism of state transfer is analyzed in comparison to the uncontrolled case. We find that the transfer evolution exhibits an asymmetry with respect to atom-field detuning, being sensitive to the sign of the detuning. Under a linear interaction controlled by the local classical fields, we show that the entanglement distribution can be both amplified and preserved against the noise. These results motivate developments towards the implementation or simulation of the purely theoretical model employing parity-deformed fields.
\end{abstract}

\date{\today}

%\pacs{42.50.Ex, 42.50.Ct, 42.50.Dv, 03.67.Bg}

\maketitle

Since the birth of quantum communication, various quantum systems
have been suggested as possible candidates for generating quantum
entanglement or transferring quantum state \cite{horodecki2009quantum,pirandolaReview,ref39,ra,
IndPRL2018,BLFC2017,BellPRA2005,quantumRepeaters,castellini2018,
sciarrinolofranco2017}. 
%Transmitting quantum states between distant sites is the goal of quantum communication, and it is one of the most challenging and rewarding tasks. 
Quantum state transfer can be completed either by 
teleportation or through quantum networking \cite{pirandolaReview,br, cl, ben1,cl1}. 
%Also, cavity quantum electrodynamics (cavity-QED) systems can serve for communicating between high-Q cavities. 
In the past two decades, many schemes of quantum communication in the context of cavity-QED have been proposed \cite{ci, enk,He2014QIP,cQED3,CQED2017,Chen2010,Guo2005,Tong2012,PhysRevLett.118.133601,song2017,beaudoin2017hamiltonian}. Most of them are based on
(in-)direct interactions between the atoms whose internal states are to be transferred. The basic idea \cite{ci, enk} is as follows: localized atoms are
trapped in high-Q cavities which are spatially separated from each
other, special laser pulses are employed and an atom emits a
photon into the cavity mode; successively, the photon enters the second
cavity and is absorbed by the other atom, so that the state transfer is
completed by appropriately switching {\em on} or {\em off} the
laser. The schemes typically utilize the interaction between
electromagnetic fields and atoms (or ions), because atoms with
long-lived internal states can be useful as storage qubits and
photons are best suited as fast information carriers.

One promising procedure consists in transmitting
the quantum state between two-level atoms located in different
cavities, with the cavity electromagnetic fields being coupled to each other. In this scenario, a robust model has been originally presented for quantum state transfer where the cavities exchange photons through a transmission line \cite{ci}. Connection between two cavities at long distance was also considered \cite{ser}, where a quantum gate is made of atoms trapped in cavities with modes that are indirectly coupled by optical fibers. 
%In recent years, schemes with these features have been proposed for entangling two or more atoms \cite{pl99} and there are essential developments in using optical fibers for quantum communication on the single photon level \cite{zb}. Based on an adiabatic passage via photonic dark states, the quantum networking with optical fibers was established \cite{pe}. The entanglement between distant atoms is prepared by generating an effective interaction of arbitrary strength between the internal degrees of freedom for the atoms placed in distant cavities connected by an optical fiber \cite{ma}. The photon leaks out of the cavities, travels through the optical fiber and the atomic system is then projected by suitable measurements into an entangled state \cite{yab}.  This shows the potential applications of optical fibers in quantum communications. 
It has been experimentally shown that entanglement can be distributed between remote parties who exchange unentangled photons \cite{fedrizzi2013experimental}. The transfer of an entangled state from atoms to cavity modes has been particularly studied \cite{zhou2006entanglement,titov2005transfer,emary2005emission,
budich2010entanglement,bougouffa2013atoms}.

On the other hand, state transfer procedures in cavity-QED networks have to face the problems due to the interaction of leaky cavities with the surrounding environment. The action of the environment on the dynamics of entanglement usually leads to its disappearance, even at a finite time \cite{zyczkowski2001k,dodd2004disentanglement, PhysRevLett.93.140404, santos2006direct, lopez2008sudden,yu2009sudden,almeida2007environment, laurat2007heralded,barbosa2010robustness}. Many efforts have been then devoted to finding strategies for entanglement protection in open quantum systems, for example by engineering structured non-Markovian (memory-keeping) environments or by suitable control techniques \cite{lofrancoreview,ref39,ref40,ref35,ref41,ref31,PhysRevA.95.052126,agarwalPRA,Mortezapour2017,Mortezapour2017b,Mortezapour2018,
Man2015,NMexp5,darrigo2012AOP,PhysRevA.81.052330,PhysRevA.85.032318,LoFrancoNatCom,LoFrancoQIP,
lofrancoPRB,NM2,NMuse3,NMexp2,RevModPhys.86.361}.

%Engineering, controlling, and simulating quantum dynamics of composite systems is a challenging task. Nevertheless, these techniques are crucial to develop quantum technologies, preserve quantum properties and engineer transformation. In the context of cavity-QED, two-level atoms can be successfully used to generate entangled states of qubits as well as realize different quantum communication protocols \cite{vanEnk2005,ra,mon}. Practical applications require robust entangled states, which entails sufficiently long lifetime of the quantum states. However, in many cases entanglement of two-level atoms is not stable enough.  To stabilize atomic entanglement, engineered protocols can be adopted \cite{kra, nic}. In particular, stabilization methods in leaky cavities have been proposed by means of an optical white-noise field \cite{xu} or by using three level $\Lambda$-type atoms in cavities  \cite{can,bis,PhysRevA.84.064302,jinLPL,shenEPL}. In addition, entanglement transfer from optical fields to separated atoms can be achieved which crucially depends on initial correlations present in a driving field and on cavity damping rates \cite{paternostro2004complete, zou2006entanglement, PhysRevA.86.052315, casagrande2009tripartite,paternostro2009passing,yonacc2010coherent,ran2016entanglement}.

In this work, we analyze the time behavior of quantum state transfer in a cavity-QED network through the theoretical model of parity-deformed radiation fields \cite{Chaichian,PhysRev.138.B1155}. In particular, we use the so-called $R$-deformed Heisenberg extension of the cavity photons \cite{deh0}, whose peculiar effect can be interpreted as the action of local external classical fields. We investigate the entanglement distribution to distant atomic qubits embedded in separated leaky cavities which are connected by an optical lossy fiber, acting as a cavity buffer with delocalized modes. To the best of our knowledge, parity-deformed fields have not been experimentally reported while some proposals of implementation exist in the literature \cite{alderete2018simulating, alderete2017quantum, alderete2017nonclassical}. 
Other than their general interest from a purely theoretical viewpoint within the context of parafields and parastatistics \cite{PhysRevLett.65.980,PhysRev.138.B1155,Plyushchay2,Plyushchay1,govorkov,Chaichian,Buzano}, our aim here is to explore their role within a quantum information scenario. 
We indeed show that the realization of such a parity-deformed cavity system would enable a high-fidelity state transfer.

The paper is organized as follows. In Sec. \ref{Sec:ParityDeformedField} we recall the Hamiltonian associated to a parity-deformed Jaynes-Cummings model for a two-level atom (qubit) in a single-mode cavity. The overall system under consideration is then described in Sec. \ref{Sec:TheSystem}.
In Sec. \ref{Sec:MEq} we derive the relevant master equation and give the structure of the evolved density matrix of the system. The dynamics of the state transfer process is provided in Sec. \ref{Sec:DynEntTrans}. In Sec. \ref{Sec:Conc} we discuss the results.

\section{Parity-deformed field and Jaynes-Cummings model} \label{Sec:ParityDeformedField}

In this section we briefly review a parity extension of the standard Jaynes-Cummings model within the rotating-wave approximation \cite{deh0}, which will be useful for our successive analysis.

It is first convenient to define the single-mode generators $\mathfrak{a}$,
$\mathfrak{a}^{\dag}$ of a parity-deformed field in terms of the well-known photon annihilation and creation operators $a$, $a^{\dag}$, which are
\begin{equation}\label{generators}
\mathfrak{\hat{a}}=\hat{a}-\frac{\lambda}{\sqrt{2}x}\hat{R},\quad
\mathfrak{\hat{a}}^{\dag}=\hat{a}^{\dag}+\frac{\lambda}{\sqrt{2}x}\hat{R},
\end{equation}
where $\lambda$ is a real constant named Wigner deformation parameter while $\hat{R}$ is a parity operator, being Hermitian and unitary with the properties
$\hat{R}^{2}=\hat{I}$, $\hat{R}^{\dag}=\hat{R}^{-1}=\hat{R}$ ($\hat{I}$ is the identity operator). This parity operator $R$ acts in the Hilbert space of the eigenvectors (number states) as $\hat{R}\ket{k}=(-1)^k\ket{k}$, while the continuous variable $x$ is the position defined by the quadrature of the generator operators $\hat{x} = (\mathfrak{a} + \mathfrak{a}^{\dag})/\sqrt{2}$ \cite{deh0}. 
The generators $\mathfrak{\hat{a}}$, $\mathfrak{\hat{a}}^{\dag}$, together with $\hat{I}$ and $\hat{R}$, construct the so-called parity-deformed Heisenberg algebra \cite{Wigner,Yang,deh0} which exhibits significant features in quantum optics
\cite{Sage, Deh1, Deh2, Deh3, Deh4, alderete2018simulating, alderete2017quantum, alderete2017nonclassical}, and satisfy the following (anti-)commutation relations
\begin{equation}
[\mathfrak{\hat{a}},\mathfrak{\hat{a}}^{\dag}]=1+2\lambda \hat{R},
\hspace{2mm}\{\hat{R}, \mathfrak{\hat{a}}\}=\{\hat{R},
\mathfrak{\hat{a}}^{\dag}\}=0.
\end{equation}
The action of $\mathfrak{\hat{a}}$ and $\mathfrak{\hat{a}}^{\dag}$ on a number (Fock) state $\ket{k}$ ($k=0,1,2,\ldots$) is, respectively, \cite{deh0}
\begin{eqnarray}
\mathfrak{\hat{a}}\ket{2k}&=&\sqrt{2k}\ket{2k-1},\nonumber\\ \mathfrak{\hat{a}}\ket{2k+1}&=&\sqrt{2k+2\lambda+1}\ket{2k},\nonumber\\
\mathfrak{\hat{a}}^{\dag}\ket{2k}&=&\sqrt{2k+2\lambda+1}\ket{2k+1},\nonumber\\ 
\mathfrak{\hat{a}}^{\dag}\ket{2k+1}&=&\sqrt{2k+2}\ket{2k+2}.
\end{eqnarray} 

The Hamiltonian of the parity-deformed Jaynes-Cummings model has therefore the form
\begin{equation}\label{Hlambda0}
H_{\lambda}:=\frac{\omega_{0}}{2}\sigma_{3}+\frac{\omega_{c}}{2}\{\mathfrak{\hat{a}},{\mathfrak{\hat{a}}}^{\dag}\}+\eta({\mathfrak{\hat{a}}}^{\dag}\sigma_{-}+\mathfrak{\hat{a}}\sigma_{+}),
\end{equation}
where $\omega_{0}$ is the atomic transition frequency, $\omega_{c}$ the cavity mode frequency, $\eta$ the atom-cavity coupling constant, while $\sigma_{\pm}$ and $\sigma_{3}$ are the usual Pauli rising (lowering) and inversion operators for the atomic
states satisfying
$[\sigma{+},\sigma_{-}]=\sigma_{3}$ and
$[\sigma_{3},\sigma{\pm}]=\pm2\sigma_{\pm}$.
With the explicit expressions of the generators given in Eq. (\ref{generators}), it can be recast into
\begin{equation}\label{Hlambda}
H_{\lambda}=H_\mathrm{JC}+\frac{\omega_{c}}{2}\frac{\lambda^2}{x^2}+\frac{\omega_{c}}{2}\left\{\hat{a}-{\hat{a}}^{\dag},\frac{\sqrt{2}\lambda}{x}\hat{R}\right\}-i\frac{\sqrt{2}\eta\lambda}{x}\hat{R}\sigma_{2},
\end{equation}
where $\sigma_2 = \sigma_y$ is the second Pauli matrix and
\begin{equation}\label{JCM}
H_\mathrm{JC}=\frac{\omega_{0}}{2}\sigma_{3}+\frac{\omega_{c}}{2}\left\{\hat{a},{\hat{a}}^{\dag}\right\}+\eta({\hat{a}}^{\dag}\sigma_{-}+\hat{a}\sigma_{+}),
\end{equation}
indicates the well-known Jaynes-Cummings Hamiltonian.
The Hamiltonian $H_{\lambda}$ can be interpreted as describing a two-level atom (qubit) coupled to a single-mode cavity field subject to a local external classical field $\hat{E}_\mathrm{ext}=-\frac{\lambda}{\sqrt{2}x}\hat{R}$ of frequency $\omega_c$ \cite{deh0}.
Such an external field can be meant as a function of the continuous variable $x$,
whose dynamical features are ruled by the parameter $\lambda$. In Eq. (\ref{Hlambda}), the second term $\frac{\omega_{c}}{2}\frac{\lambda^2}{x^2}$ thus represents the self-energy of the external field, the third term
$\frac{\omega_{c}}{2}\left\{\hat{a}-{\hat{a}}^{\dag},\frac{\sqrt{2}\eta\lambda}{x}\hat{R}\right\}$ denotes the interaction between the quantized cavity field and
an external field and, finally,
$-i\frac{\sqrt{2}\eta\lambda}{x}\hat{R}\sigma_{2}\equiv \hat{E}_\mathrm{ext}d(i\sigma_2)$ identifies the coupling between the atom and an external classical field, with $d$ recalling the atomic dipole matrix element for the transition.

Notice that the standard Jaynes-Cummings model is retrieved by using the usual annihilation and creation operators $a$, $a^\dag$ in place of the parity-deformed ones. In fact, when $\lambda = 0$ the Hamiltonian $H_{\lambda}$ of Eq. (\ref{Hlambda}) reduces to $H_\mathrm{JC}$.

\section{The system} \label{Sec:TheSystem}

\begin{figure}[t!]
\begin{center}
\includegraphics[width=0.5\textwidth]{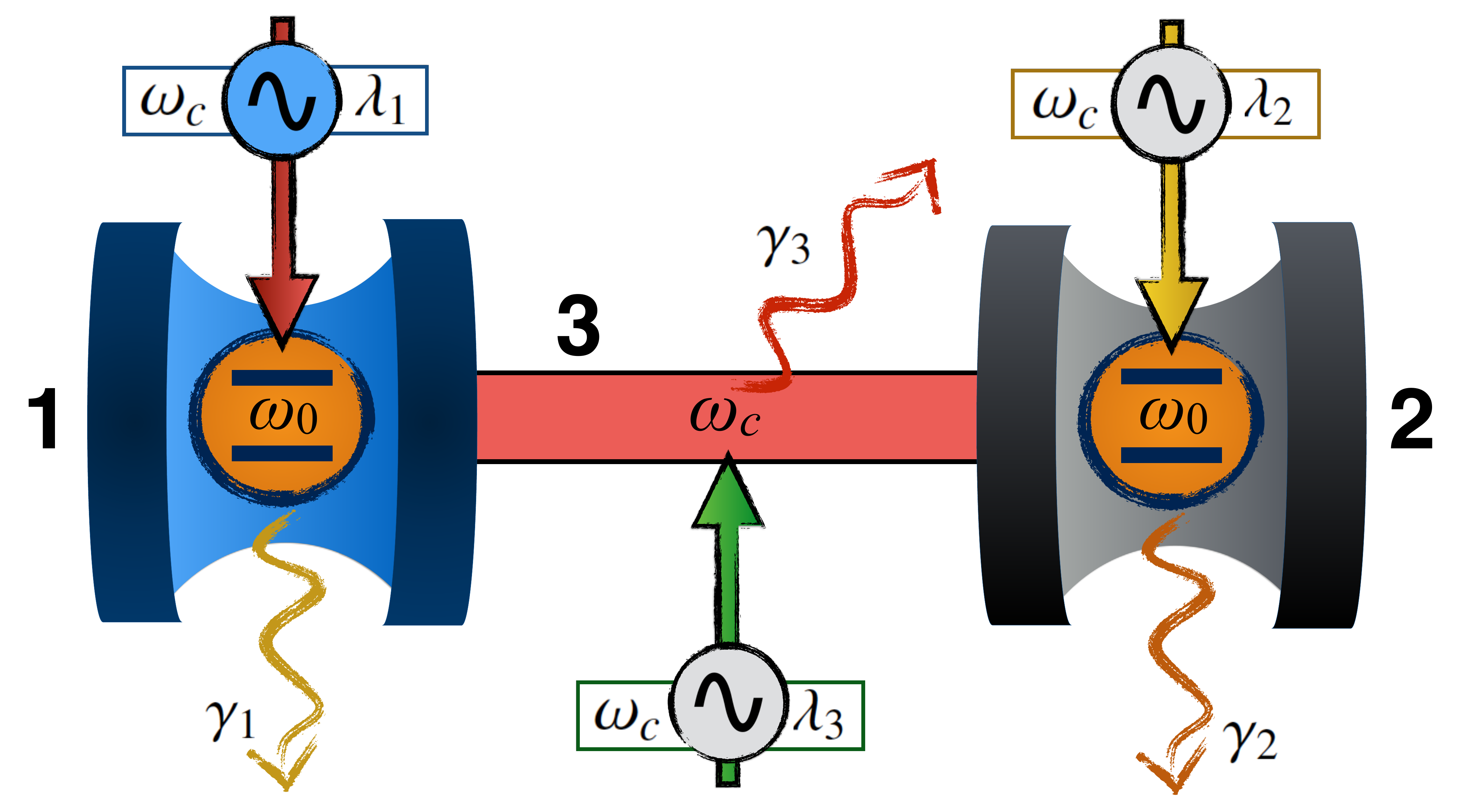}
\end{center}
\caption{\textbf{Sketch of the system.} Two qubits of transition frequency $\omega_0$ are embedded in two separated cavities linked via a single-mode fiber (cavity buffer). Cavities (1 and 2) and fiber (3) have frequency $\omega_c$ and leak photons to the surrounding bath at rates $\gamma_1$, $\gamma_2$ and $\gamma_3$, respectively. The three cavity radiation fields are parity-deformed ones, whose effect is intepretable as qubits, cavities and fiber all driven by local classical fields with frequency $\omega_c$ and amplitudes ruled by the Wigner parameters $\lambda_i$ ($i=1,2,3$).} \label{Fig1}
\end{figure}

We now consider the composite system of interest for our study, which is made of two separated qubits (two-level atoms) each one embedded in a leaky cavity and governed by the parity-deformed Jaynes-Cummings model described above \cite{deh0}, the two cavities being connected by a lossy optical fiber, as depicted in Fig. \ref{Fig1}. The fiber is a cavity buffer also described by a parity-deformed field to simulate its interaction with a local classical field. All the subsystems (atoms, cavities and fiber) are open and surrounded by an external bath at thermal equilibrium (zero temperature).

The evolution of the total system (including the bath) is ruled by the Hamiltonian
\begin{equation}\label{Hamilt}
H = H_\mathrm{s}+H_\mathrm{bath}+H_\mathrm{int},
\end{equation}
where $H_\mathrm{s}$ corresponds to the Hamiltonian of the system composed of
the atoms, cavities and fiber, $H_\mathrm{bath}$ is the bath Hamiltonian and $H_\mathrm{int}$ is the system-bath interaction Hamiltonian.

The system Hamiltonian describes all the relevant interactions among the various subsystems, with atoms, cavities and fiber subject to local classical fields. It has the form
\begin{equation}\label{Hs}
H_\mathrm{s}
=\sum^{2}_{i=1}H_{{\lambda_{i}}}+\frac{\omega_{c}}{2}\{\mathfrak{\hat{a}}_{3},\mathfrak{\hat{a}}_{3}^{\dag}\}+\upsilon
\mathfrak{\hat{a}}_{3}(\mathfrak{\hat{a}}_{1}^{\dag}+
\mathfrak{\hat{a}}_{2}^{\dag})+\upsilon\mathfrak{\hat{a}}_{3}^{\dag}(\mathfrak{\hat{a}}_{1}+
\mathfrak{\hat{a}}_{2}),
\end{equation}
where $\mathfrak{\hat{a}}_{i}$ are the boson-like generator operators
for the first and second cavity ($i=1, 2$) and for the optical fiber ($i = 3$), whose expression is that of Eq. (\ref{generators}). The only difference among these
operators for the three parts (cavity-1, cavity-2, fiber) is in the possible
different values of the real parameters $\lambda_{i}$ ($i=1,2,3$) and the continuous
variables $x$, in general.
In Eq. (\ref{Hs}), each of the terms $H_{{\lambda_{i}}}$ represents a parity-deformed Jaynes-Cummings Hamiltonian as given in Eq. (\ref{Hlambda0}), the second term
$\frac{\omega_{c}}{2}\{\mathfrak{\hat{a}}_{3},\mathfrak{\hat{a}}_{3}^{\dag}\}$
denotes a free Hamiltonian of the fiber mode including the interaction term between the fiber itself and a local external field, and the last
terms $\upsilon\mathfrak{\hat{a}}_{3}(\mathfrak{\hat{a}}_{1}^{\dag}+
\mathfrak{\hat{a}}_{2}^{\dag})+\upsilon\mathfrak{\hat{a}}_{3}^{\dag}(\mathfrak{\hat{a}}_{1}+\mathfrak{\hat{a}}_{2})$ describe the interaction
between parity-deformed fields. Notice that $\omega_{c}$ is the fiber mode frequency, which is the same of that of the cavity modes and of the local classical fields, while $\upsilon$ denotes the cavity-fiber hopping strength.

Indicating with $b_{ij}$ $({b_{ij}}^{\dag})$ the usual bath annihilation (creation) operators for the three subsystems ($i=1,2,3$), the bath and interaction Hamiltonians appearing in Eq. (\ref{Hamilt}) are, respectively, $H_\mathrm{bath}=\sum^{\infty}_{j=1}\sum^{3}_{i=1}{\omega_{ij}{b_{ij}}^{\dag}b_{ij}}$ and
\begin{equation}
H_\mathrm{int} =
\sum^{\infty}_{j=1}\sum^{3}_{i=1}{\Omega_{ij}\left(\mathfrak{\hat{a}}_{i}+
\mathfrak{\hat{a}}_{i}^{\dag}\right)\left(\hat{b}_{ij}+\hat{b}_{ij}^{\dag}\right)},
\end{equation}
where $\Omega_{ij}$ are the system-bath coupling constants.

Since we are interested in the dynamics of entanglement transfer involving qubits and cavities at zero temperature, we assume that only a single excitation is at most contained within the system. Indicating with $\ket{g}$ and $\ket{e}$ the ground and excited states of each atom qubit, the bare basis of the system can be denoted by $\{|n\rangle,\ n =1,\ldots,6\}$, with $|1\rangle:=|eg000\rangle$,
$|2\rangle:=|gg100\rangle$, $|3\rangle:= |gg001\rangle$, $|4\rangle:=
|gg010\rangle$, $|5\rangle:= |ge000\rangle$ and the ground (zero-excitation) state $|6\rangle:= |gg000\rangle$. Notice that, in this notation, the order of each subsystem state in the system state vector is $\ket{\textrm{qubit-1}\ \textrm{qubit-2}\ \textrm{cavity-1}\ \textrm{cavity-2}\ \textrm{fiber}}$.

\section{Master equation and evolved density matrix} \label{Sec:MEq}

To describe how the quantum state transfer works, we need the evolved density matrix of the system. To this aim, we firstly introduce the (dressed) eigenstates $|\phi_{n}\rangle$ ($n=1,\ldots,6$) corresponding to each eigenvalue $\epsilon_{n}$ of the system Hamiltonian $H_\mathrm{s}$ which for $n=1,\ldots,5$ are
\begin{equation}\label{phin}
|\phi_{n}\rangle = c_{n1}|1\rangle+c_{n2}|2\rangle+c_{n3}|3\rangle
+c_{n4}|4\rangle+c_{n5}|5\rangle,
\end{equation}
while $|\phi_{6}\rangle=|6\rangle$ is the ground state. We point out that ${\mathfrak{\hat{a}}}_{i}^{\dag}|\phi_{n}\rangle =0$ ($i=1,2,3$; $n=1,\ldots,5$)  because the number of excitations cannot be larger than one, while ${\mathfrak{\hat{a}}}_{i}^{\dag}|\phi_{6}\rangle\neq0$. The explicit expressions of the coefficients $c_{ni}$ of Eq. (\ref{phin}) are rather cumbersome and are thus not reported here: their values shall be calculated numerically for given values of the system parameters.

To obtain the evolved density matrix of the system, we use the standard Liouville-von Neumann equation \cite{ref30} for the total density operator in the interaction picture with respect to $H_\mathrm{s} + H_\mathrm{bath}$. Performing the Born-Markov and rotating-wave approximation, tracing out the environmental (bath) degrees of freedom and then going back to the Schr\"{o}dinger picture, one obtains the microscopic master equation \cite{scala2007,Scala2007b,Gonz2018} for the reduced density operator $\rho(t)$ of the system at zero temperature as
\begin{equation}\label{MasterEq}
\dot{\rho}(t) = -i[H_\mathrm{s},
\rho(t)]+\sum^{3}_{i=1}\sum_{n=1}^5\gamma_{i}\left[{\hat{A}}_{i,n}\rho(t)
{\hat{A}}_{i,n}^{\dag}-\frac{1}{2}\left\{{\hat{A}}_{i,n}^{\dag}{\hat{A}}_{i,n},\rho(t)\right\}\right],
\end{equation}
where $\gamma_i$ ($i=1,2,3$) are the photon decay rates for the cavities and fiber, respectively, and
\begin{equation}\label{Aop}
\hat{A}_{i,n} = |\phi_{6}\rangle\langle
\phi_{6}|\left(\mathfrak{\hat{a}}_{i}+
\mathfrak{\hat{a}}_{i}^{\dag}\right)|\phi_{n}\rangle\langle
\phi_{n}|.\quad (n=1,2,\ldots,5)
\end{equation}
At zero temperature the system can make transitions only
downwards on the energy ladder. For this reason, the operators of
Eq. (\ref{Aop}) become
$\hat{A}_{i,n}= |\phi_{6}\rangle\langle
\phi_{6}|\mathfrak{\hat{a}}_{i}|\phi_{n}\rangle\langle \phi_{n}|$,
which give
\begin{eqnarray}
\hat{A}_{1,n} &=&
c_{n2}\sqrt{2\lambda_{1}+1}|\phi_{6}\rangle\langle \phi_n|,\nonumber\\
\hat{A}_{2,n}&=&
c_{n4}\sqrt{2\lambda_{2}+1}|\phi_{6}\rangle\langle \phi_n|,\nonumber\\
\hat{A}_{3,n} &=&
c_{n3}\sqrt{2\lambda_{3}+1}|\phi_{6}\rangle\langle \phi_n|.
\end{eqnarray}

Using the master equation of Eq. (\ref{MasterEq}) and the notation
${\rho}(t)_{mn}=\langle \phi_{m}|{\rho}(t)|\phi_{n}\rangle$ ($m,n=1,\ldots,5$), the differential equations for the elements of the system reduced density matrix
$\hat{\rho}(t)$ are
\begin{eqnarray}\label{DiffEq}
&\dot{\rho}(t)_{nn}=-\gamma_{nn}{\rho}(t)_{nn},\quad 
\dot{\rho}(t)_{66}=\sum^{5}_{n=1}{\gamma_{nn}{\rho}(t)_{nn}},&\nonumber\\
&\dot{\rho}(t)_{nm}=\left[i(\epsilon_{m}-\epsilon_{n})-\frac{\gamma_{nn}+\gamma_{mm}}{2}\right]{\rho}(t)_{nm},\quad
\dot{\rho}(t)_{6n}=0,&
\end{eqnarray}
where $\gamma_{nn}:=\gamma_{1}(2\lambda_{1}+1)|c_{n2}|^2+\gamma_{2}(2\lambda_{2}+1)|c_{n4}|^2+\gamma_{3}(2\lambda_{3}+1)|c_{n3}|^2$.
Since the evolution of the system is written in the
dressed-state basis $\{|\phi_n\rangle\}$, it is strategical to perform a change of basis
\begin{equation}\label{basischange}
|n\rangle =
\tilde{c}_{n1}|\phi_{1}\rangle+\tilde{c}_{n2}|\phi_{2}\rangle+\tilde{c}_{n3}|\phi_{3}\rangle
+\tilde{c}_{4n}|\phi_{4}\rangle+\tilde{c}_{n5}|\phi_{5}\rangle,
\end{equation}
where, we have defined the matrix $\tilde{C}= \{\tilde{c}_{mn}\}$ by inverting the matrix of the above mentioned coefficients $C= \{c_{mn}\}$ (see Eq. (\ref{phin})), i.e. $\tilde{C}=C^{-1}$. Then, one can easily determine the solutions of the first-order differential equations in Eq. (\ref{DiffEq}) as
\begin{eqnarray}
&\rho(t)_{nn}={\rho}(0)_{nn}e^{-\gamma_{nn}t},\quad
\rho(t)_{66}=\sum^{5}_{n=1}{{\rho}(0)_{nn}\left[1-e^{-\gamma_{nn}t}\right]},&\nonumber\\
&\rho(t)_{nm}={\rho}(0)_{nm}e^{\frac{2i(\epsilon_{m}-\epsilon_{n})-(\gamma_{nn}+\gamma_{mm})}{2}t},\quad \rho(t)_{6n}=\rho(0)_{6n}.&
\end{eqnarray}

The general expressions found above for the evolved system density matrix are a very convenient starting point for our analysis. In fact, by choosing the system initial state ${\rho}(0)$ and the values of the various parameters, they allow us to obtain the evolved reduced density matrix of the subsystems of interest. In particular, since we are interested in transferring entanglement between the two separated qubits, we shall focus on the two-qubit evolved density matrix.

\section{Dynamics of entangled state transfer} \label{Sec:DynEntTrans}

In this section, we show how the above parity-deformed Hamiltonian and the corresponding evolved density matrix of the system can be used for investigating the transfer of entanglement to atomic qubits.

An arbitrary state of the system at time $t$, expressed in the dressed state basis $\{\ket{\phi_n}, n=1,\ldots,6\}$, can be written as
\begin{eqnarray}
{\rho}(t) &=& \sum^{6}_{m,n=1}{{\rho}(t)_{mn}|\phi_{m}\rangle\langle
\phi_{n}|}.
\end{eqnarray}
Tracing over cavities and fiber degrees of freedom, we get the two-qubit reduced density matrix
\begin{eqnarray}\label{twoqubitDM}
\varrho(t) &=& \varrho(t)_{1}|eg\rangle\langle
eg|+\varrho(t)_{2}|ge\rangle\langle ge|+
\varrho(t)_{3}|gg\rangle\langle gg|\nonumber\\
&+&\varrho(t)_{4}|eg\rangle\langle ge|+
\bar{\varrho}(t)_{4}|ge\rangle\langle eg|,
\end{eqnarray}
where the coefficients are given by
\begin{eqnarray}
\varrho(t)_{1} &=&
\sum^{5}_{n=1}{{\rho}(t)_{nn}|c_{n1}|^2+2\sum^{5}_{n\neq m=1}{\mathfrak{Re}[{\rho}(t)_{nm}\overline{c_{m1}}c_{n1}]}},\nonumber\\
\varrho(t)_{2} &=&
\sum^{5}_{n=1}{{\rho}(t)_{nn}|c_{n5}|^2+2\sum^{5}_{n\neq m=1}{\mathfrak{Re}[{\rho}(t)_{nm}\overline{c_{m5}}c_{n5}]}},\nonumber\\
\varrho(t)_{3} &=&
{\rho}(t)_{66}+\sum^{5}_{n=1}\sum^{4}_{i=2}{\rho(t)_{nn}|c_{ni}|^2}\nonumber
\\
&&+2\sum^{5}_{n=1}\sum^{4}_{i, m=2}{\mathfrak{Re}\left(\rho(t)_{nm}c_{ni}\overline{c_{mi}}\right)},\nonumber\\
\varrho(t)_{4} &=&
\sum^{5}_{n=1}{{\rho}(t)_{nn}\overline{c_{n5}}c_{n1}+\sum^{5}_{n\neq
m=1}{{\rho}(t)_{nm}\overline{c_{m5}}c_{n1}}},
\end{eqnarray}
with $\overline{z}$ indicating the complex conjugate of a number $z$. Such equations permit us to study the desired entanglement evolution, once chosen the initial state $\rho(0)$ of the system.

In the standard two-qubit product computational basis $\{|ee\rangle , |eg\rangle ,
|ge\rangle , |gg\rangle\}$, we evaluate the entanglement which builds up between the two atom qubits during the evolution using the well-known concurrence measure \cite{horodecki2009quantum}. For the evolved two-qubit density matrix $\varrho(t)$ of Eq. (\ref{twoqubitDM}), the concurrence at time $t$ is $C(\varrho(t))=2\max\{0, |\varrho(t)_{4}|\}$.

\subsection{Entanglement transfer from cavities to atoms}

\begin{figure*}[t!]
\begin{center}
\includegraphics[width=0.7\textwidth]{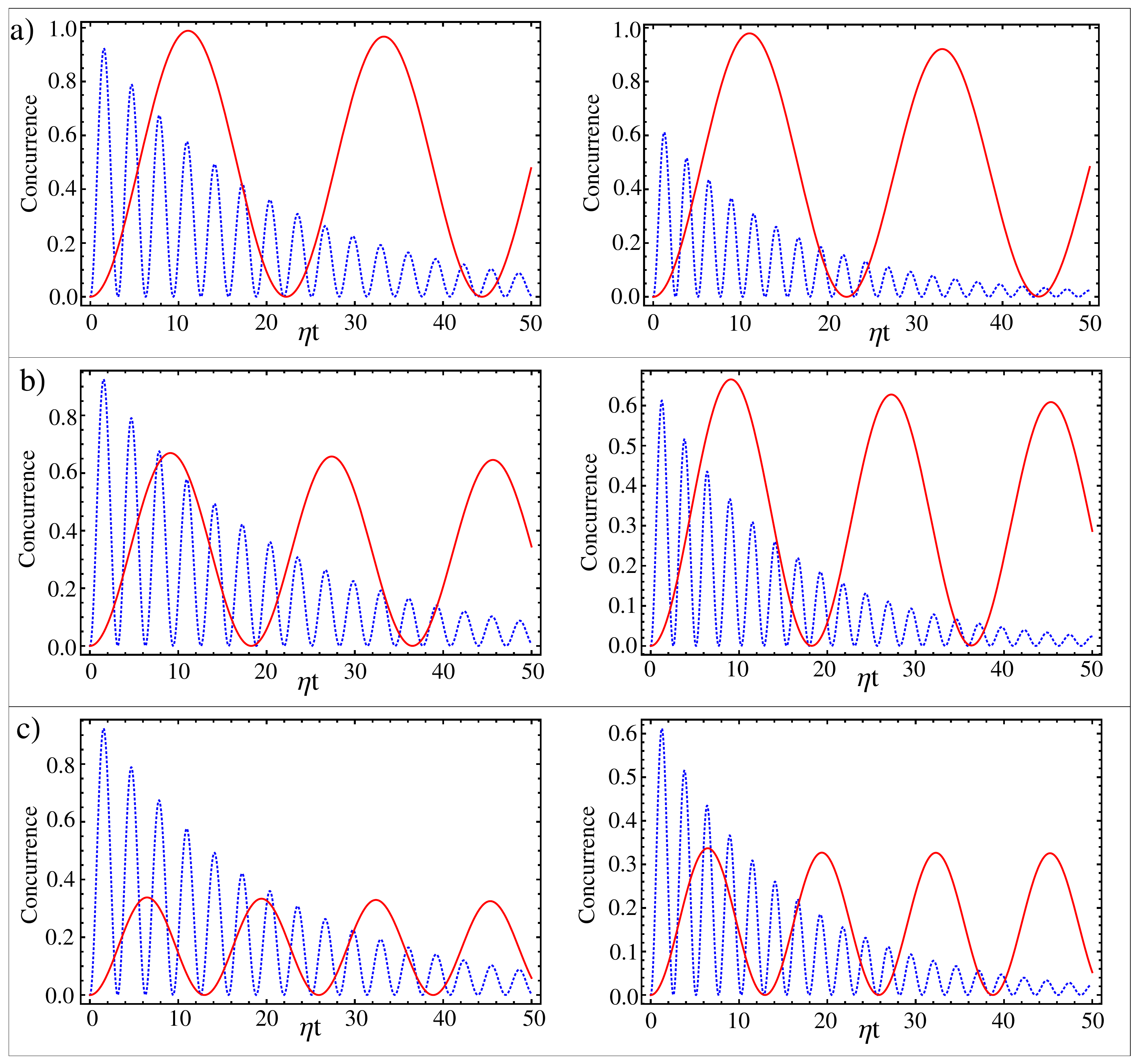}
\end{center}
\caption{\textbf{Entanglement transfer from cavities to atoms.} Dynamics of two-qubit entanglement (concurrence) as a function of the dimensionless time $\eta t$. The curves are plotted for the set of parameters: $\omega_{0}=0.2\eta$, $\lambda_{1}= \lambda_{2}=\lambda_{3}= -0.49$, $\gamma_{1}= \gamma_{2}= \gamma_{3}= 0.1\eta$ and $\upsilon= 0.5\eta$ (plots are independent of the value of $x$). The plots display the results for the standard (uncontrolled) Jaynes-Cummings Hamiltonian (blue dashed line) and for the parity-deformed (controlled) Hamiltonian (red solid line). The plots are related to the initial system states $\ket{\Psi^{-}}$ (left plots) and $\ket{\Psi^{+}}$ (right plots). Panels (a), (b) and (c) correspond to different detuning $\Delta=-0.1\eta$, $0$ and $0.1\eta$, respectively.} \label{Fig2}
\end{figure*}

We first study the transfer of entanglement from the cavity modes to the atom qubits.
We take at the initial time the fiber mode in the vacuum state $\ket{0}$ and the atoms in the ground state $|gg\rangle$, while the two cavities are in one of the two Bell states $\ket{\psi^{\pm}}=(\ket{10}\pm\ket{01})/\sqrt{2}$. Notice that such an entangled state for the two cavity modes can be prepared by standard cavity-QED or circuit-QED procedures by suitably controlling atom-cavity interactions \cite{ra,vanEnk2005,Devoret1169}.

The system initial state is therefore $\rho(0) =\ket{\Psi^{\pm}}\bra{\Psi^{\pm}}$ with
\begin{equation}
|\Psi^{\pm}\rangle=
\left(\frac{|gg100\rangle\pm|gg010\rangle}{\sqrt{2}}\right)=\frac{|2\rangle\pm|4\rangle}{\sqrt{2}},
\end{equation}
where we have used the bare state basis $\{\ket{n}\}$ of the system. Passing to the dressed basis $\{\ket{\phi_n}\}$ by means of Eq. (\ref{basischange}), one then obtains the initial density matrix elements
\begin{eqnarray}
{\rho}(0)_{nn}&=&\frac{|\tilde{c}_{2n}+\tilde{c}_{4n}|^2}{2},\quad
{\rho}(0)_{nm}=\frac{(\tilde{c}_{2n}+\tilde{c}_{4n})(\overline{\tilde{c}_{2m}}+\overline{\tilde{c}_{4m}})}{2},\nonumber\\
\rho(0)_{6n}&=&0.
\end{eqnarray}

We are now able to numerically analyze and plot the evolution of entanglement transfer from the two cavities to the two atoms by choosing suitable values of the system parameters. In general, the dynamics is sensitive to the different values of all the involved parameters (except the continuous variables $x$, which does not affect the dynamics). In Fig. \ref{Fig2} we plot the concurrence as a function of the dimensionless time $\eta t$, fixing suitable values of the parameters $\omega_0$, $\lambda_i$, $\gamma_i$ and $\upsilon$, comparing it with the uncontrolled case without local classical fields given by the standard Jaynes-Cummings interaction ($\lambda_i=0$). We observe that, despite similar qualitative periodic behaviors with the amplitudes eventually decreasing in time, the dynamics of transfer shows quantitative differences depending on the initial state and, especially, on the atom-field detuning $\Delta=\omega_{c}-\omega_{0}$. Compared to the decreasing oscillations occurring for the standard Jaynes-Cummings interaction, the control of local classical fields in the parity-deformed Jaynes-Cummings model has the general effect to improve the stabilization and protection of the two-qubit entanglement established during the evolution. The best performance is found starting from the initial state
$\ket{\Psi^{+}}$ (right plots of Fig. \ref{Fig2}) and out of resonance with negative detuning $\Delta$ (red-detuned field). It is interesting to point out the asymmetry shown by the concurrence as regards the detuning. In fact, one could reckon that the performance of state transfer is symmetric with respect to the resonant case, independently of the sign of the field detuning $\Delta$. Instead, we observe that negative atom-field detuning ($\Delta=-0.1\eta$) enables a more efficient entanglement transfer with respect to the resonant case, while positive detuning ($\Delta=0.1\eta$, blue-detuned field) supplies a smaller efficiency. Differently, one immediately sees that the uncontrolled dynamics is not significantly affected by the detuning.
We mention that analogous results would be obtained for the inverse entanglement transfer process, that is from qubits to cavities, as expected due to the interchangeability of each local qubit and cavity mode.

\begin{figure*}[t!]
\begin{center}
\includegraphics[width=0.7\textwidth]{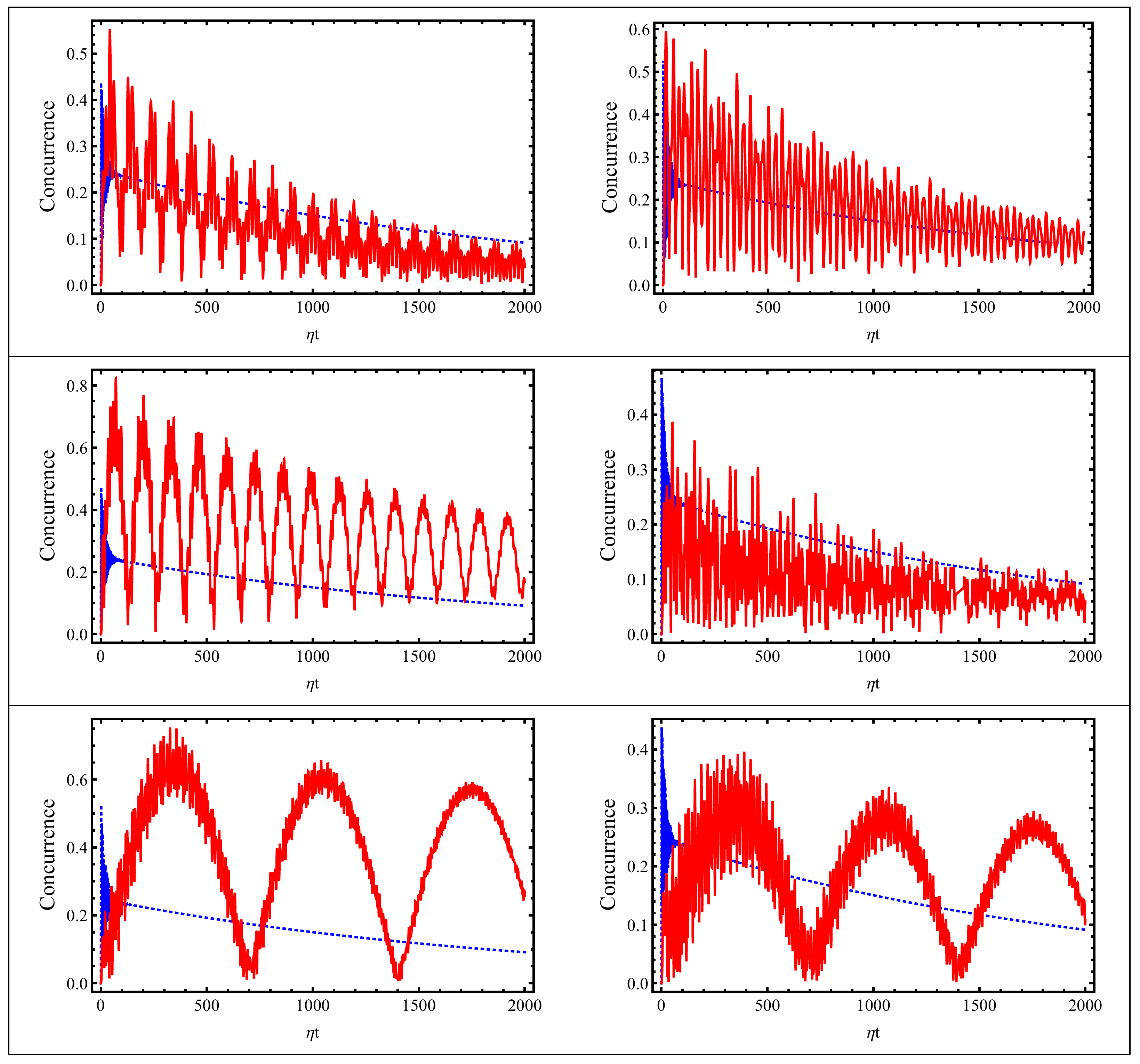}
\end{center}
\caption{\textbf{Entanglement transfer from a local atom-cavity subsystem to separated atoms.} Dynamics of two-qubit entanglement (concurrence) as a function of the dimensionless time $\eta t$. The curves are plotted for the set of parameters: $\omega_{0}=0.4\eta$, $\lambda_{1}= \lambda_{2}=\lambda_{3}= -0.49$,
$\gamma_{1}= \gamma_{2}= \gamma_{3}= 0.1\eta$ and
$\upsilon= 10\eta$ (plots are independent of the value of $x$). The plots display the concurrence resulting from the standard (uncontrolled) Jaynes-Cummings Hamiltonian (dashed line) and from the parity-deformed (controlled) Hamiltonian (solid line). The plots are related to the initial system states $\ket{\Phi^{-}}$ (left plots) and $\ket{\Phi^{+}}$ (right plots). Panels (a), (b) and (c) are related to different values of detuning $\Delta=-0.2\eta$, 0 and $0.2\eta$, respectively. } \label{Fig3}
\end{figure*}

\subsection{Entanglement transfer from a local atom-cavity subsystem to separated atoms}

We here investigate the state transfer performance from a local hybrid atom-cavity entanglement, for instance initially created between qubit-1 and cavity-1, to the two separated atom qubits in the subsystems 1 and 2. We then choose fiber, cavity-2 and qubit-2 initially in their vacuum and ground states, respectively, with qubit-1 and cavity-1 being instead in one of the two Bell states $\ket{\phi^{\pm}}=(\ket{e0}\pm\ket{g1})/\sqrt{2}$. Such an initial local atom-cavity entangled state can be as well prepared by usual cavity-QED (or circuit-QED) interaction-based methods \cite{ra,vanEnk2005,Devoret1169}.

The global initial state of the system is $\rho(0)=\ket{\Phi^\pm}\bra{\Phi^\pm}$ with
\begin{equation}
|\Phi^\pm\rangle=\frac{|eg000\rangle\pm|gg100\rangle}{\sqrt{2}}=\frac{|1\rangle\pm|2\rangle}{\sqrt{2}},
\end{equation}
which leads to the following initial density matrix elements in the dressed state basis
\begin{eqnarray}
{\rho}(0)_{nn}&=&\frac{|\tilde{c}_{2n}+\tilde{c}_{1n}|^2}{2},\quad
{\rho}(0)_{nm}=\frac{(\tilde{c}_{2n}+\tilde{c}_{1n})(\overline{\tilde{c}_{2m}}+\overline{\tilde{c}_{1m}})}{2},\nonumber\\
\rho(0)_{6n}&=&0.
\end{eqnarray}

The time behavior of the concurrence associated to the two-atom evolved density matrix corresponding to this initial condition is plotted in Fig. \ref{Fig3}, for some fixed values of the system parameters. It is worth mentioning that the chosen values for the parameters have been used in the numerical analysis to obtain particularly interesting performances for the entanglement distribution from both the qualitative and quantitative viewpoints. 
%We first notice that an almost stationary entanglement occurs for the uncontrolled case governed by the standard Jaynes-Cummings interactions in each local subsystem.The formation of such a stabilized entangled state may be explained by the role cavity-1 assumes in this case of initial hybrid qubit-cavity entangled state, in analogy to the stationary entanglement occurring for two qubits embedded in a common cavity \cite{ref65}. Cavity-1, through the (open) fiber, can be in fact thought as a catalyst of the entanglement for the nonlocal hybrid (qubit-1)-(cavity-2) system, with cavity-2 which in turn exchanges excitations with the atom qubit-2. 
We look for a good amplification and shield from noise for the entanglement transfer, which is shown to happen for the controlled case of the parity-deformed Jaynes-Cummings interaction.

The optimal performance is reached starting from the state $\ket{\Phi^-}$ and for the resonant case, as seen in the left plot of Fig. \ref{Fig3}(b), where the concurrence attains its higher value (larger than $0.8$) which is periodically retrieved. At variance with the process of entanglement transfer from cavities to qubits treated above, the resonant case provides here a high fidelity entanglement transfer to the qubits. An asymmetric behavior as regards the detuning $\Delta$ is still retrieved, both qualitatively and quantitatively. In particular, a more chaotic and slower amplification of the state transfer is found for positive detuning ($\Delta=0.2\eta$) with respect to negative detuning ($\Delta=-0.2\eta$). In general the action of the local classical fields, although weakening the stabilization, remains strategic for amplifying the efficiency of the process of entanglement distribution.

\section{Conclusions} \label{Sec:Conc}

In this work we have studied the effects of the theoretical model of parity-deformed radiation fields into the efficiency of entanglement distribution to distant qubits in a noisy cavity network. We have considered the simple situation of two noninteracting atom qubits embedded in separated leaky cavities which are in turn linked by a dissipative fiber, which functions as a cavity buffer with delocalized modes.
The parity-deformed cavity photons introduce additional terms to the system Hamiltonian which can be seen as external classical fields locally controlling each atom-cavity subsystem as well as the fiber.

The dynamics of entanglement transfer depends on both the initial state of the system and the detuning between the qubit transition frequency and the cavity field frequency. The (parity-deformed) process starting from initially entangled cavities to the qubits is found to be more stable with respect to that starting from a local atom-cavity entangled state. The larger gain, however, is achieved in the latter situation, where the parity-deformed (controlled) model allows an amount of entanglement transfer up to four times larger than that obtained for the (uncontrolled) case ruled by a standard Jaynes-Cumming interaction. 
%Moreover, preparing a local atom-cavity entanglement may be less demanding than creating an initial entangled state for the separated cavities of the network.
Depending on the initial state of the system, the resonant condition corresponding to zero atom-cavity detuning does not always provide the most efficient process, whose dynamics is quite sensitive to the sign of the detuning. This asymmetry as regards the detuning is a peculiar dynamical aspect, since one may expect that the evolution of state transfer is symmetric with respect to the resonant case, independently of the sign of the atom-field detuning.

As a general result, we have shown that the adoption of parity-deformed cavity fields, introducing the action of specific local classical fields as external control, can enable robust transfer of entangled states to far qubits of a cavity-based quantum network. In particular, the entanglement distribution can be both amplified and preserved against the noise compared to the case without control.
%The system is conceptually simple and is in principle suited for scalability by a straightforward extension to a many-node cavity network, where two adjacent cavities are connected by a fiber acting as a cavity buffer. 
Albeit the model of parity-deformed cavity radiation here employed remains a purely theoretical one, our results highlight its potentiality in a quantum information scenario, possibly motivating developments towards the implementation or simulation of such parafields in contexts like trapped atoms in cavity-QED or artificial atoms in circuit-QED \cite{ra,vanEnk2005,Devoret1169,kimbleReview}.

%\bibliography{ref_EntTrans}

\end{document}